\documentclass[12pt,a4paper]{article}
\usepackage{indentfirst}
\usepackage[utf8]{inputenc}
\usepackage[english]{babel}
\usepackage{mflogo}
\usepackage{amsmath,amsfonts,amssymb}
\usepackage{graphicx}
\usepackage{cite}

\usepackage{amsthm,amsmath}

\theoremstyle{plain}

\newtheorem*{theorem*}{Theorem}

\newtheorem*{lemma*}{Lemma}
\newtheorem*{remark}{Remark}

\author{ M. V. Karasev\thanks{e-mail:  \texttt{mkarasev@hse.ru}}
 \qquad  E. V. Vybornyi\thanks{e-mail: \texttt{evybornyi@hse.ru}} \\ \textit{
National Research University Higher School of Economics, Moscow, Russia.}
}
\title{Tunnel catch from potential wells and energy detection}
\date{\vspace{-5ex}}

\begin{document}
\maketitle
\begin{abstract}
We consider the one-dimensional Schr\"{o}dinger operator in the semiclassical regime assuming that its double-well potential is the sum of a finite ``physically given'' well and a square shape probing well whose width or depth can be varied (tuned). We study the dynamics of initial state localized in the physical well. It is shown that if the probing well is not too close to the physical one and if its parameters are specially tuned, then the {\it tunnel catch effect} appears, i.e. the initial state starts tunneling oscillations between the physical and probing wells. The asymptotic formula for the probability of finding the state in the probing well is obtained. We show that the observation of the tunnel catch effect can be used to determine the energy level of the initial state, and we obtain the corresponding asymptotic formula for the initial state energy. We also calculate the leading term of the tunneling  splitting of energy levels in this double well potential.
\end{abstract}

\section{Introduction}

The quantum tunneling in a double-well potential is a core phenomenon~\cite{MR, JA}. The first qualitative results about the spectrum were already obtained back in 1927~\cite{FH}.
The analytic description of this phenomenon can be achieved in the semiclassical regime, i.~e. asymptotically by ``Plank constant'' $\hbar\to0$. The semiclassical approximation for the quantum tunneling is well studied in the case of a mirror-symmetric double-well potential. Namely the first semiclassical formulas for tunneling energy splitting was obtained in~\cite{L} for the case of high exited states, and in~\cite{Slav} for the case of lower energy levels, also see~\cite{F, Harrell1, Harrell2, ADS, GP} and therein for more details. 
The multidimensional generalization of the asymptotic formula for tunneling splitting was rigorously proved in~\cite{DKM} for the pair of ground state energies, also see~\cite{Herring, BDS, DK, BS1}.
Some deep results about semiclassical spectrum and the corresponding stationary states (as ``avoided crossing'' and ``the flea on the elephant'' effects) are also known in the case of asymmetric double-well potential~\cite{EV, JL, P2}, including the multidimensional case~\cite{HS1,HS2,HS3,BS,BS2}. 

Recall that if the double-well potential is symmetric, the spectrum of the Schr\"{o}dinger operator consists of pairs of exponentially close (as $\hbar\to0$) points and the corresponding stationary states are symmetric and antisymmetric. 
If a state is asymptotically localized in only one well then it is not a stationary state, but it can be represented as a linear combination of symmetric and antisymmetric stationary states with slightly different energies. This results in the effect of tunneling transport when a state initially localized in one well starts tunneling oscillations between two wells with a certain period. The period depends on the tunnel splitting of the energy levels.

Similar effect can also take place for an asymmetric double-well potential under a priori condition that stationary states are localized in both potential wells~\cite{Bender}. The general criterion of such a bilocalization of stationary states for asymmetric double-well potentials was obtained in~\cite{EV}.

In the given paper we consider a specifically chosen double-well potential that is equal to the sum of a basic ``physically given'' well and a square shape probing well. Let us assume that the physical well is a generic smooth finite function
\footnote{If the actual physical potential is not finite but it rapidly tends to zero faraway from the well one can introduce an appropriate effective finite well and take into account the discarded non-finite part of the potential as a small perturbation.}.
To be definite, assume that the probing well is placed on the right from the physical well (see Fig. 1). The parameters of the probing square well (depth, width and position) are varied.

We study the dynamics of a state initially localized in the physical well.
If parameters of the probing well are generic then the property of localization in the physical well is preserved in time (with exponential accuracy as $\hbar\to0$). But for some specific values of parameters the situation is changed: the initial state is leaving the physical well, transported to the probing one, then returns back to the physical well, and so on. The states starts oscillations between two wells. Thus under a specific tuning of parameters the observer can detect the initial physical state via its full appearance in the probing well. This phenomenon can be referred as \textit{tunnel catch} effect.
Based on the results of~\cite{EV} we derive the explicit analytic conditions for appearance of this effect  (see Eq.~(\ref{eqCondSq}),~(\ref{eqw})).

Note that the general problem to find conditions for the tunnel catch are interesting from physical standpoint~\cite{TD}; for particular examples they were considered  in~\cite{D} using numerical methods.

We show that the observation of the tunnel catch effect can be used to detect the energy level of the physical well (see Conclusions).
We also derive an asymptotic formula for the probability of finding the state in the probing well (see Eq.~(\ref{eqPmax})) and calculate the leading term of the tunneling splitting of energy levels in our specific double well potential.
\begin{figure}
\center{\includegraphics[scale=0.2]{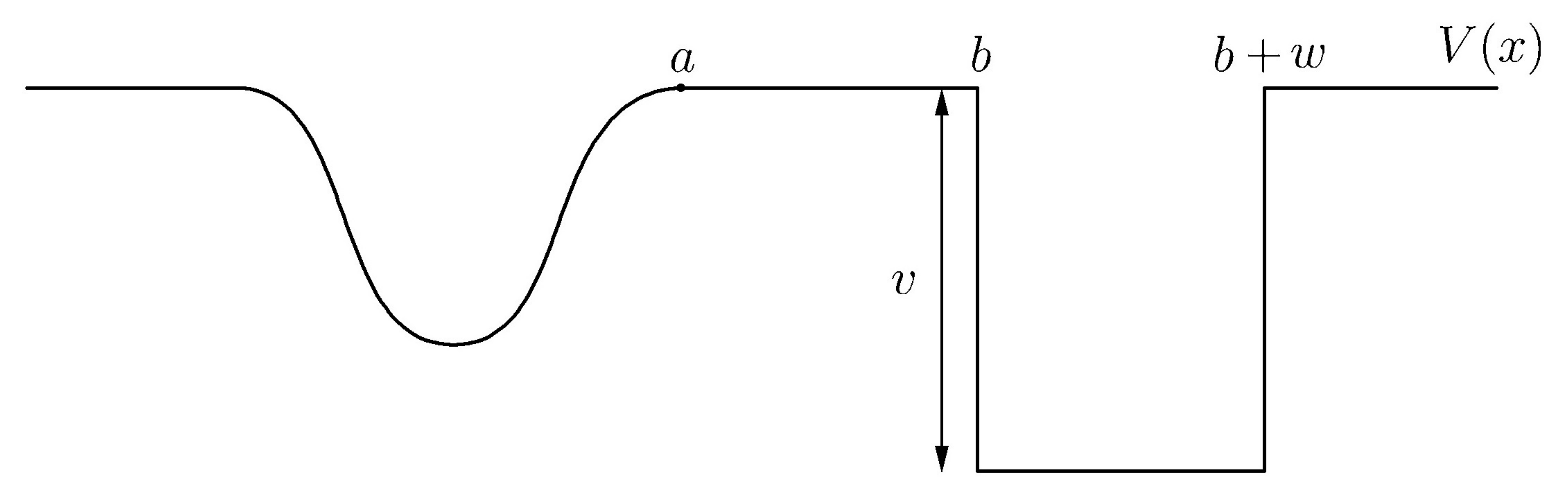}}
\center{Fig. 1}
\end{figure}

\section{Statement of the problem and basic definitions}
\label{sec2}
We consider the one-dimensional Schr\"{o}dinger equation
\begin{equation}
\label{eqSh}
\left\{
\begin{aligned}
&i\hbar\frac{\partial \Psi}{\partial t}=\hat H \Psi,\\
&\Psi\bigl|_{t=0}=\Psi_0,
\end{aligned}
\right.
\end{equation}
where $\hbar>0$ is the small parameter of semiclassical approximation and the Schr\"{o}dinger operator is
$$\hat H=-\hbar^2 \frac{d^2}{dx ^2}+V(x).$$
The potential $V(x)$ is equal to the sum of two negative finite functions whose supports do not intersect,
$$V(x)=V_l(x)+V_r(x),$$
$$V_i(x)\le 0,\ i=l,r.$$
The physical (left) well is smooth (see Fig. 1),
$$V_l(x)=0,\ x\ge a,$$
and the probing (right) well has a square shape
$$V_r(x)=\left\{
\begin{aligned}
&0 &&\ x\le b,\ x\ge b+w;\\
&-v &&\ b<x<b+w.
\end{aligned}
\right.
$$
The width $w$ of the probing well is a controllable parameter.

We assume that the initial state $\Psi_0$ is localized in the physical well with energy  close to a negative value  $E>-v$. For simplicity, let $\Psi_0$ be just equal to the eigenstate $\psi_l$ corresponding an eigenvalue $E=E_l$ of the Schr\"{o}dinger operator with potential $V_l(x)$. Thus the ``physical'' initial state $\Psi_0=\psi_l$ is completely independent from the probing well.

The probabilities $ P_ {l} (t) $ and $ P_r (t) $ of finding the state $\Psi$ in the left and in the right potential wells at time $ t $ can be used to detect chances in the state localization. The probabilities are calculated as integrals of $|\Psi(x,t)|^2$ by $x$ over neighborhoods of the left and right wells respectively. The maximal probability of finding the state in the probing well is
$$P_r^{max}=\max_t P_r(t).$$
The {\it tunnel catch effect} is said to appear if  $P_r^{max}$ does not tend to zero as $\hbar\to0$. We show below that $P_r^{max}$ approaches $1$ as $\hbar\to0$ for certain discrete values of the parameter $w$. Thus the full size tunnel catch effect appear, if the probing well is specially tuned.

\section{Two level approximation}
\label{sec3}
Let us consider a generic double-well potential that is the sum of two finite potential wells. The corresponding Schr\"{o}dinger operators is
$$\hat H=-\hbar^2\frac{d^2}{dx^2}+V(x),$$
where
$$V(x)=V_l(x)+V_r(x).$$
It is well known that the spectrum of Schr\"{o}dinger operator with double well potential could contain pairs of exponentially close points and corresponding eigenstates could be localized in both wells~\cite{L, Harrell1, Harrell2}.
Using results of~\cite{EV}, we derive below asymptotic formulas for spectrum of $\hat H$ and for corresponding eigenstates.

First let us introduce two Schr\"{o}dinger operators
$$\hat H_l=-\hbar^2 \frac{d^2}{dx ^2}+V_l(x),$$
$$\hat H_r=-\hbar^2 \frac{d^2}{dx ^2}+V_r(x).$$
The operators $\hat H_{l,r}$ separately describe the left and right finite wells. 
Let $x_l$ and $x_r$ be the left and right turning points of the potential barrier (see. Fig. 2).

\begin{figure}
\center{\includegraphics[scale=0.2]{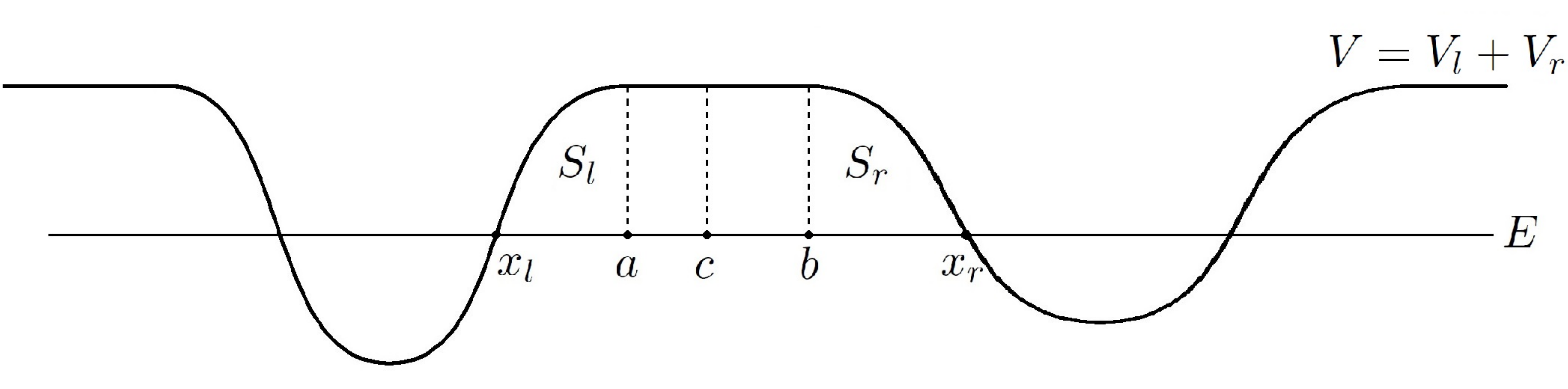}}
\center{Fig. 2}
\end{figure}

Consider a point $c$ such that
$$\int_{x_l}^{c}|p(x)|dx=\int_{c}^{x_r}|p(x)|dx,$$
where $p(x)=\sqrt{E-V(x)}$. The point $c$ is a {\it potential barrier center} from the standpoint of the action in the instanton sense. The leading role of instanton metric in double-well potentials was studied in~\cite{BS, BS2, Agmon,Maslov} and in series of papers~\cite{HS1, HS2, HS3}.

It was shown in~\cite{EV} that the asymptotic formulas for the spectrum and eigenstates of $\hat H$ could be presented in terms of the spectrum of $\hat H_l$ and $\hat H_r$ if
\begin{equation}
\label{eqCond}
a<c<b.
\end{equation}
This inequality means that the point $c$ does not belong to the support of $V(x)$. It is equivalent to the inequality
$$
\left|S_r-S_l\right|<(b-a)\sqrt{-E},\eqno \text{(\ref{eqCond}a)}
$$
where $S_{l,r}$ are tunnel actions on the left and right banks of the barrier,
$$S_l=\int_{x_l}^a\sqrt{V_l(x)-E}dx,\quad S_r=\int_{b}^{x_r}\sqrt{V_r(x)-E}dx.$$

Let $E_{i}$ be an eigenvalue of $\hat H_{i}$ and $\psi_i(x)$ be the corresponding eigenstate, $i=l, r$.
Then the spectrum of $\hat H$ near energy $E$ consists of points $E_l$ and $E_r$ with exponential accuracy~\cite{EV} as $\hbar\to0$. The eigenstate of $\hat H$ either tends to $\psi_l$ or $\psi_r$ if the distance between $E_l$ and $E_r$ is not exponentially small. Conversely, if it is exponentially small, then there is a pair of exponentially close points in the spectrum of $\hat H$, and the corresponding eigenstates tend to linear combinations of $\psi_l$ and $\psi_r$.

\begin{lemma*}
\label{lemm}
Let the condition~(\ref{eqCond}a) holds and the energies $E_l$ and $E_r$ coincide with each other:
$$E_l=E_r=E.$$
Then there are two points $E_1$ and $E_2$ in spectrum of $\hat H$ exponentially close to $E$ as $\hbar\to0$. The corresponding eigenstates of $\hat H$ are localized in both potential wells and have the form
\begin{equation}
\label{eqpsi12simpl}
\begin{split}
&\psi_{1}\simeq\frac{1}{\sqrt{2}}(\psi_l + \psi_r ),\\
&\psi_2\simeq\frac{1}{\sqrt{2}}(\psi_l - \psi_r ),
\end{split}
\end{equation}
with exponential accuracy as $\hbar\to0$.
\end{lemma*}

Moreover, if $E_l\ne E_r$, but the distance between $E_l$ and $E_r$ is exponentially small as $\hbar\to0$, then the points $E_{1,2}$ of the spectrum of $\hat H$ satisfies the following approximate formula:
\begin{equation}
\label{eqE12}
E_{1,2}\simeq\frac{E_r+E_l}{2}\pm\frac{1}{2}\sqrt{\delta^2+(E_r-E_l)^2},
\end{equation}
where
\begin{equation}
\label{eqdpsi}
\delta=2\hbar^2\left[\psi_l\frac{d\psi_r}{dx}-\psi_r\frac{d\psi_l}{dx}\right]_{x=c}
\end{equation}
Therefore the leading term of the splitting $\Delta=E_2-E_1$ is
\begin{equation}
\label{eqD}
\Delta\simeq\sqrt{\delta^2+(E_r-E_l)^2}
\end{equation}
The corresponding eigenstates of $\hat H$ have the form
\begin{equation}
\label{eqpsi12}
\begin{split}
&\psi_1\simeq\psi_l \cos{\alpha}+\psi_r \sin{\alpha},\\
&\psi_2\simeq\psi_l \sin{\alpha}-\psi_r\cos{\alpha},
\end{split}
\end{equation}
where
\begin{equation}
\label{eqtan}
\tan{\alpha}=\frac{E_r-E_l}{\delta} + \sqrt{1+\frac{(E_r-E_l)^2}{\delta^2}},
\end{equation}

It follows from~(\ref{eqtan}) that the eigenstates $\psi_{1,2}$ are localized in both the left and right potential wells if $|E_r-E_l|=O(\delta(\hbar))$. Conversely, the eigenstates $\psi_{1,2}$ are exponentially close to $\psi_{l,r}$ if $|E_r-E_l|$ is much greater than $\delta(\hbar)$.

Lemma formulated above and equations~(\ref{eqE12})-(\ref{eqtan}) follow from results of~\cite{EV} that were derived based on the two-level approximation method.

\begin{remark}
We do not formulate here the exact condition on smoothness of the potential wells, but it certainly includes the case of piecewise continuous functions, which is discussed in the next section.
\end{remark}

\section{The square shape probing well}
\begin{figure}
\center{\includegraphics[scale=0.8]{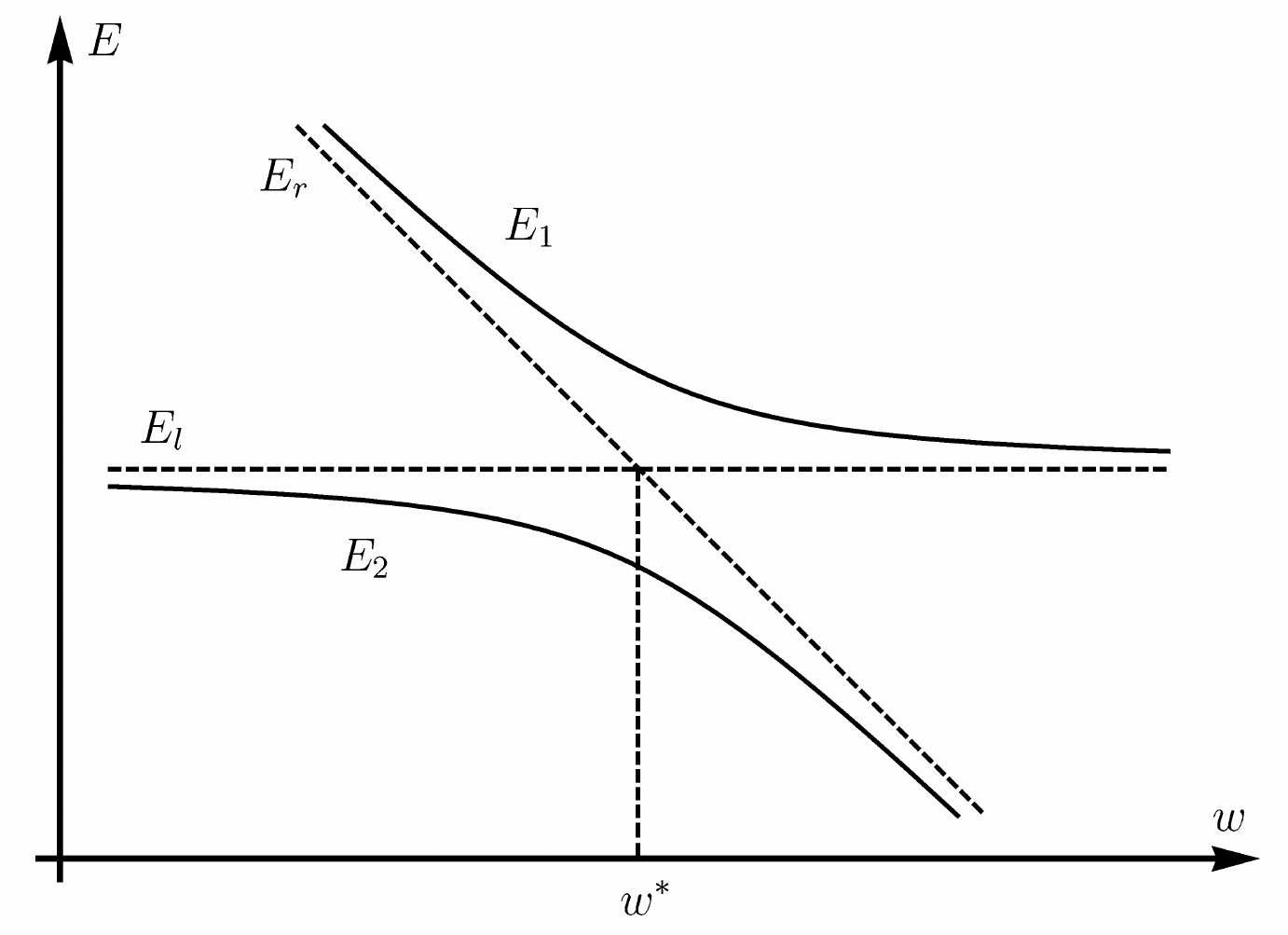}}
\center{Fig. 3}
\end{figure}

Let the probing (right) well be of square shape, as it was described in Section~\ref{sec2}. Then condition~(\ref{eqCond}a) becomes
\begin{equation}
\label{eqCondSq}
b-a>\int\sqrt{1-V_l(x)/E}\, dx.
\end{equation}
The integral in~(\ref{eqCondSq}) is taken over the interval $[x_l,a]$ where the physical well $V_l$ varies from the value $E<0$ to the value $0$.
We suppose the condition~(\ref{eqCondSq}) is satisfied, i.e. the probing well is far enough from the physical well.

The negative discrete energy levels $E=E_r$ of the square shape probing well are the solutions of the following equation
\begin{equation}
\label{eqEr}
\frac{w\sqrt{v+E}}{\hbar}=
\pi(k+1/2) - \arctan{\left(
\frac{v+2E}
{2\sqrt{(-E)(v+E)}}\right)}.
\end{equation}
Here $k\ge0$ be the number of the energy level $E=E_r^{(k)}$. Thus the distance between adjacent energy levels of the operator  $\hat H_r$ has order $\hbar^2$ near the bottom of the square potential well and has order $\hbar$ for high energy levels  with $k\sim 1/\hbar$.

Using explicit equation~(\ref{eqEr}), we can easily obtain approximate formulas for $E_r^{(k)}$. Namely the asymptotic expansion of $E_r^{(k)}$ as $\hbar\to0$ for finite $k\ge0$ is
\begin{equation}
\label{eqErk}
E_r^{(k)}= -v +\hbar^2 \frac{\pi^2 (k+1)^2}{w^2}\left(1 - \frac{4}{w\sqrt{v}}\hbar + \frac{12}{w^2 v} \hbar^2 -\frac{2(48+(k+1)^2\pi^2)}{3v^{3/2}w^3}\hbar^3+ \ldots\right).
\end{equation}

By substituting the WKB approximation of $\psi_l$ and the explicit form of $\psi_r$ into~(\ref{eqdpsi}), we derive
\begin{equation}
\label{eqd}
\delta=2\hbar (-E)^{1/4} \sqrt{\frac{ 2 (v+E) \omega_l}{\pi v w}} \exp{\left(-\frac{1}{\hbar}\int_{x_l}^{x_r}|p|dx\right)}\left[1+O(\hbar)\right],
\end{equation}
where $\omega_l$ is the classical frequency of oscillations in the left well $V_l(x)$ for the energy $E$.

Let the width $w=w_k^*$ of the probing well be tuned in such a way that the energy $E_l$ of the initial state $\psi_l$ coincides with the energy $E_r^{(k)}$ of the probing well (see Fig. 3):
$$E_l=E_r^{(k)}\big|_{w=w^*_k}.$$
From~(\ref{eqEr}) we compute
\begin{equation}
\label{eqw}
w_k^*=\frac{\pi\hbar}{\sqrt{v+E_l}}\left[k+\frac{1}{2} - \frac{1}{\pi}
\arctan{\left(\frac{v+2E_l}{2\sqrt{(-E_l)(v+E_l)}}\right)}\right].
\end{equation}

\begin{theorem*}
Let condition~(\ref{eqCondSq}) holds and the width of the probing well $w$ coincide with one of the resonance value $w^*_k$~(\ref{eqw}). Then the initial state localized in the probing well starts tunneling oscillations between wells, and the maximal probability of finding the state in the probing well approaches $1$ as $\hbar\to0$.
\end{theorem*}

\begin{remark}
We can similarly consider the depth $v$ of the probing will as a varying parameter, and formulate a similar theorem. Therefore, there is the series of the resonance values $v=v_k^{*}$ such that the energy of the initial state $E_l$ coincide with one of the energies $E_r^{(k)}$ of the probing well. In contrast to explicit formula~(\ref{eqw}) for the resonance width values, using equation~(\ref{eqEr}), we can obtain only an approximate formula for the resonance depth values $v_k^{*}$ 
\begin{equation}
\label{eqvk}
v_k^{*}=-E_l+\hbar^2\frac{\pi^2(k+1)^2}{w^2}\left(
1 - \frac{4}{w\sqrt{-E_l}}\hbar - \frac{12}{w^2 E_l}\hbar^2 - \frac{4(24-(k+1)^2\pi^2)}{3w^3 (-E_l)^{3/2}}\hbar^3+\ldots\right),
\end{equation}
where $k\ge0$ is a finite number.
\end{remark}

\begin{proof}
Using the lemma of Sect.~\ref{sec3}, we expand the initial state $\Psi_0=\psi_l$ with respect to eigenstates of operator~$\hat H$:
$$\psi_l=\frac{1}{\sqrt{2}}(\psi_1+\psi_2).$$
Then the solution of the Cauchy's problem~(\ref{eqSh}) with exponential accuracy as $\hbar\to0$ is
$$\Psi(x,t)=e^\frac{t(E_1+E_2)}{2 i\hbar}\left(
\cos\left(t\frac{E_2-E_1}{2\hbar}\right)\psi_l+
i\sin\left(t\frac{E_2-E_1}{2\hbar}\right)\psi_r\right),$$
where $E_2-E_1\simeq\delta$. It is clear that the state $\Psi$ is oscillating between the physical well and the probing well, and the state $\Psi$ is localized only in the probing will with exponential accuracy when $t=\frac{\pi\hbar}{E_2-E_1}$.
\end{proof}

Thus we obtain the series of width values $w=w^*_k$ such that the fixed energy level $E_l$ of the physical well coincides with levels $E_r^{(k)}$ of the probing well. For these values of the parameter $w$ the probability $P_r^{max}$ approaches $1$, see Fig.~4. Thus if the probing well is well tuned, namely if its width $w$ coincides with one of the values $w_k^*$~(\ref{eqw}) then the probing well ``catch'' the state with the energy $E_l$ localized in the physical well. 

Further, let us consider the width of the probing well $w$ to be close to the critical value $w^*_k$ (see Fig. 3 and 4) to determine the width of resonance peaks of $P_r^{max}$. Using~(\ref{eqpsi12}) we derive
$$\Psi_0=\psi_1 \cos{\alpha}+\psi_2\sin{\alpha}.$$
Then the asymptotic solution of the Cauchy's problem~(\ref{eqSh}) is
$$\Psi=\left(e^{\frac{t E_1}{i\hbar}}\cos^2\alpha+e^{\frac{t E_2}{i\hbar}}\sin^2\alpha\right)\psi_l+
\cos{\alpha}\sin{\alpha}\left(e^{\frac{t E_2}{i\hbar}}-e^{\frac{t E_1}{i\hbar}}\right)\psi_r.$$

\begin{figure}
\center{\includegraphics[scale=0.18]{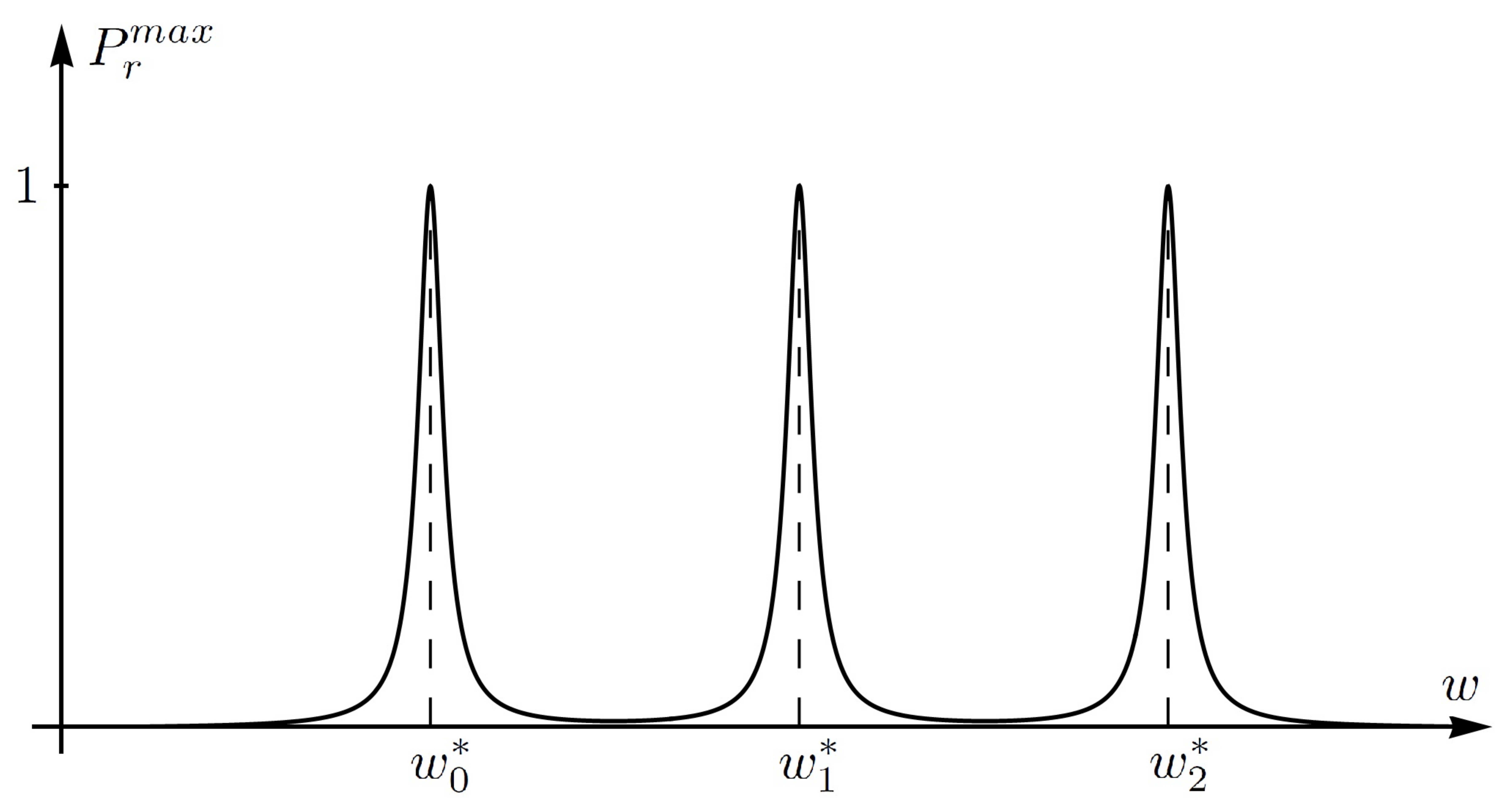}}
\center{Fig. 4}
\end{figure}

The probability $P_{l}(t)$ and $P_r(t)$ of finding the state $\Psi$ in the left and right potential wells at time $t$ are
$$P_l(t)=\int_{-\infty}^{c}|\Psi(x,t)|^2dx,$$
$$P_r(t)=\int_{c}^{\infty}|\Psi(x,t)|^2dx.$$
Therefore, we derive
$$P_l(t)=\cos^4\alpha+\sin^4\alpha+2\cos^2\alpha\sin^2\alpha\cos\left(\frac{E_2-E_1}{\hbar}t\right),$$
$$P_r(t)=2\cos^2\alpha\sin^2\alpha\left(1-\cos\left(\frac{E_2-E_1}{\hbar}t\right)\right),$$
$$P^{max}_{r}=\max\limits_{t}P_r(t)=4\cos^2\alpha\sin^2\alpha.$$
Using~(\ref{eqtan}), we finally obtain
\begin{equation}
\label{eqPmax}
P^{max}_{r}=\frac{\delta^2}{\delta^2+(E_r-E_l)^2}=\left(\frac{\delta}{\Delta}\right)^2.
\end{equation}
Thus, the state significantly manifest itself in the probing well if and only if $|E_r-E_l|= O (\delta)$. The width of resonance peaks (see Fig. 4) is exponentially small and has oder $O(\delta)$ as $\hbar\to0$.

\section{Conclusion}
We obtain the explicit formula for resonance values of the external parameter~(\ref{eqw}) and for the width of the resonance peaks~(\ref{eqPmax}) in the case of the double-well potential presented at Fig. 1. We also obtain approximate formula~(\ref{eqvk}) for resonance values of the probing well depth $v$ if it is considered as a varying parameter instead of the width $w$.

These results can be used to determine the energy of the initial state localized in the physical potential well (the left in Fig. 1). Namely, let us adiabatically increase the width $w$ (or the depth $v$) of the probing (right) well, starting from such a small value that there is no bounded state in the probing well. Thus, it is possible to determine the first resonance value of the external parameter (first resonance peak in Fig. 4) and find the corresponding energy of the initial state by the formula~(\ref{eqErk}) with $k=0$
$$E=-v+\frac{\pi^2}{w^2}\hbar^2 - \frac{4\pi^2}{w^3\sqrt{v}}\hbar^3+\ldots$$
A further increase of the varying parameter will lead to the disappearance of the tunneling catch effect, and then again to its appearance at the next resonance value. Therefore, we can determine the energy level of the general physical well by the detection of the tunnel catch effect varying a parameter of the probing well.

The tunneling catch of the state initially localized in one well is a general effect. For example, one can similarly obtain it by consider a different configuration of the probing will. This effect owes to an avoided crossing of energy levels in the case of general one-dimensional double-well potential depending on an external varying parameter.

\section*{Acknowledgments}
This work was done at the Chair of Applied Mathematics under support of the Academic Foundation of the National Research University Higher School of
Economics.

\end{document}